\documentclass[epj]{svjour}
\usepackage{graphics}
\usepackage{epsfig}
\usepackage{amssymb}
\usepackage{amsfonts}
\usepackage{here}
\begin{document}
  \title{Invariant mass distributions for the $pp\to pp\eta$ reaction at Q~=~10~MeV}
%
\author{
 P.~Moskal\inst{1}\inst{,2} \and
 R.~Czy\.zykiewicz\inst{1}\inst{,3} \and
 E.~Czerwi\'nski\inst{1}\inst{,2} \and
 D.~Gil\inst{1} \and
 D.~Grzonka\inst{2} \and
 L.~Jarczyk\inst{1} \and
 B.~Kamys\inst{1} \and
 A.~Khoukaz\inst{4} \and
 J.~Klaja\inst{1}\inst{,2} \and
 P.~Klaja\inst{1}\inst{,2} \and
 W.~Krzemie{\'n}\inst{1}\inst{,2} \and
 W.~Oelert\inst{2} \and
 J.~Ritman\inst{2} \and
 T.~Sefzick\inst{2} \and
 M.~Siemaszko\inst{3} \and
 M.~Silarski\inst{1} \and
 J.~Smyrski\inst{1} \and
 A.~T\"aschner\inst{4} \and
 M.~Wolke\inst{5} \and
 P.~W\"ustner\inst{2} \and
 J.~Zdebik\inst{1} \and
 M.~J.~Zieli\'nski\inst{1} \and 
 W.~Zipper\inst{4} 
}                     
\institute{
Institute of Physics, Jagellonian University, PL-30-059 Cracow, Poland
 \and
Nuclear Physics Institute, Research Center J{\"u}lich, D-52425 J\"{u}lich, Germany
 \and
Institute of Physics, University of Silesia, PL-40-007 Katowice, Poland
 \and
IKP, Westf\"alische Wilhelms-Universit\"at, D-48149 M\"unster, Germany
 \and
Department of Physics and Astronomy, Uppsala University, Sweden
}
\date{Received: date / Revised version: date}
%
\abstract{Proton-proton and proton-$\eta$ invariant mass 
distributions and the total cross section 
for the $pp\to pp\eta$ reaction
have been determined near the threshold at an excess energy of Q~=~10~MeV.
The experiment has been conducted using the COSY-11 
detector setup and the cooler synchrotron COSY.
The determined invariant mass spectra reveal significant
enhancements in the region of low proton-proton relative momenta, 
similarly as observed previously at higher excess energies of Q~=~15.5~MeV and Q~=~40~MeV.
} 
\maketitle

\section{Introduction}

The complexity of the structure of hadrons
constitutes the basic difficulty in the quantitative description 
of the hadronic interaction in the medium energy regime.
Therefore, this interaction is not well understood especially 
in the 
meson-nucleon and meson-meson 
sector,
where an additional difficulty is the relatively poor experimental data-base.
Particularly challenging are investigations
of interactions involving flavour-neutral mesons. 
This is due to the short life-time of these mesons which 
can neither be used as targets nor as beams. 
Thus, in practice such interactions can be accessed only indirectly via  observables like 
excitation functions or invariant 
mass distributions. Measurements of these observables are especially  useful
in the close-to-threshold region where the final state particles are produced with 
low relative velocities.
Among the basic flavour neutral mesons the $\eta$ is of particular 
interest since its interaction with nucleons is strong enough to be detectable
with the presently achievable experimental precision~\cite{hab},
and since its interaction seems to be sufficently 
strong to form an eta-mesic nucleus~\cite{haider257,li515}. 
The existence of such kind of nuclear matter is vividly 
discussed~\cite{boundtheory}
and there are ongoing experimental programs searching for a signal 
of such a state~\cite{bound-exp}.

The earlier high statistics measurements of the $pp\to pp\eta$ 
reaction at an excess energy of 
Q~=~15.5~MeV from the COSY-11 collaboration~\cite{pawel-prc}, and also 
the measurements of the TOF group~\cite{tof}  at Q~=~15 and 41~MeV, revealed 
that there exist significant enhancements in the invariant mass distributions
of $pp$ and $p\eta$ subsystems
at higher values of proton-proton invariant mass   
and lower values of the proton-$\eta$ invariant mass. 
One of the plausible explanations for these enhancements could be an influence
of the proton-$\eta$ interaction~\cite{hab,fix}. 
If this is the case one could use such observables
for the estimation of the strength of this interaction. However, the observed 
invariant mass distributions
could be also plausibly explained by contributions 
of higher partial waves~\cite{kanzo-acta23,kanzo045201}
or by an energy dependence of the primary production amplitude~\cite{deloff,ceci1}.
Therefore, in order to verify the 
correctness of the 
proposed explanations 
it is of importance to investigate the dependence of the strength of the enhancements
as a function of the excess energy. 
Qualitatively, with decreasing excess energy the contribution
from the higher partial waves should decrease whereas the influence 
of the interaction should be more pronounced.

Certainly,  most effectively,  contributions from higher partial waves could be
disentangled by the determination of the analysing powers 
and spin transition coefficients~\cite{kanzo045201,meyer064002,hanhartreview},
yet such investigations are not planned in the near future at COSY which is at present 
the only laboratory where it can be conducted. 
This makes 
the determination of the energy dependence
of the 
distributions of the 
differential cross section 
for the $pp\to pp\eta$ reaction
even more important for  studies 
of the proton-$\eta$ hadronic interaction
and for studies of the properties of nucleon resonances~\cite{kanzo-acta23,kanzo0803}.

In this article we present 
distributions of the proton-proton and proton-$\eta$ invariant masses at the excess energy 
of Q~=~10~MeV which is significantly closer to the threshold with respect to the previous studies. 
Although the original experiment at  Q~=~10~MeV has been devoted to the 
investigations of the analysing power for the $pp\to pp\eta$ reaction~\cite{czyzyk}
and has been performed with a polarised proton beam, the data enable also the determination
of the spin averaged observables after appropriate 
offline "depolarisation" of the beam explained in 
section~\ref{depol}. In Section~\ref{exp} we briefly describe the experimental set-up,
present the experimental principles of the measurement, and describe the method of the data
analysis. In section~\ref{results} the determined spectra 
are compared to the analogoues results determined at the excess energy of Q~=~15.5~MeV,
and the final conclusions are drawn.

\section{Experiment}
\label{exp}
\begin{figure}[H]
\begin{center}
  \includegraphics[width=7cm]{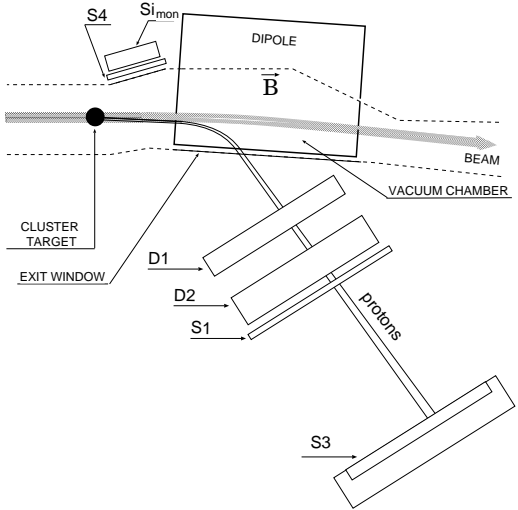}
\end{center}
\caption{ Schematic view of the COSY-11 detector setup~\cite{cosy11}.
D1 and D2 represent drift chambers.
S1, S3, S4 denote scintillator counters. 
Si$_{mon}$ is  
the silicon strip detector used for the detection 
of the elastically scattered protons. 
Superimposed solid lines indicate final state protons 
from the $pp \to pp\eta$ reaction. 
The size of detectors and their relative distances are not to scale.
\label{cosy11}
}
\end{figure}

The measurement of the $pp\to pp\eta$ reaction has been performed 
at the cooler synchrotron COSY~\cite{cosy} at the Research Center J\"ulich in Germany
using the COSY-11 detector setup~\cite{cosy11}, 
presented schematically in Figure~\ref{cosy11}.
The proton beam with a momentum
of  2.010~GeV/c, corresponding to an excess energy
of Q~=~10~MeV, has been scattered on H$_{2}$ molecules from 
an internal cluster jet target~\cite{dombrowski,khoukaz}, installed 
in front of the COSY magnet. Reaction products carry lower momenta
than the beam protons, therefore are bent more in the magnetic field 
of the dipole magnet. Positively charged ejectiles leave the scattering 
chamber through a thin exit window reaching the detection system 
operating under atmospheric pressure. 
The hardware trigger was based on signals from scintillator detectors only.
It was adjusted to register all events with at least two positively charged
particles. For this aim coincident signals in  the S1 and S3 detectors 
were required. In the case of the S1 detector only these events were
accepted for which 
either two separate segments were hitted or 
an amplitude of the signal in a single module was higher than the certain threshold value.
Based on the data analysis from the previous experiments, 
the threshold was set high enough to reduce significantly the number of
single particle events, and at the same time to accept most events (almost 100\%)
with two protons passing through one segment.
Next, in the off-line analysis it was required that at least two tracks are reconstructed 
from signals measured 
by means of two planar drift chambers D1 and D2.
The trajectories of the positively 
charged particles reconstructed in drift chambers
are further traced through the magnetic field of the dipole back to the 
interaction point. 
In this way the momenta of the particles can be reconstructed 
with a precision of 6~MeV/c (standard deviation)~\cite{pawel-prc}.
The time-of-flight measurement between the scintillator hodoscope S1 and 
the scintillator wall S3, 
and the independently reconstructed momentum  enable a particle
identification by means of the  invariant mass technique. 
The COSY-11 mass resolution allows for a clear seperation of groups
of events with  two protons, two pions, proton and pion and also deutron and pion~\cite{hab}.
Further on the produced meson is identified using the missing mass method.
A more detailed description of the method and results of the identification 
of the $pp\to pp\eta$ reaction 
can be found in reference~\cite{hab,pawel-prc,czyzyk-phd}.

\subsection{Off-line depolarisation of the beam}
\label{depol}
Originally the experiment was dedicated to the measurement
of the beam analysing
power for the $pp\to pp\eta$ reaction~\cite{czyzyk,czyzyk-phd}. 
Therefore, the proton beam 
has been vertically polarised. 
The vertical polarisation of the beam 
is defined as an asymmetry of  
populations of particles in the spin up ($N_{+}$) and down ($N_{-}$) states with respect to the vertical
axis,
integrated over the whole period of measurement:
\begin{equation}
P_y = \frac{N_+ - N_-}{N_+ + N_-}.
\label{polar}
\end{equation}  
In the discussed experiment the direction of the polarisation was being 
flipped from cycle to cycle. 
Hence, for the so called "spin up cycles" 
we define the spin up polarisation as:
\begin{equation}
P^{\uparrow} = \frac{\sum_{i=up} n_{+,i} - \sum_{i=up} n_{-,i}}{\sum_{i=up} n_{+,i} + \sum_{i=up} n_{-,i}},
\label{spin_up}
\end{equation}
and analogously for "spin down cycles" the spin down polarisation reads:
\begin{equation}
P^{\downarrow} = \frac{\sum_{i=dn} n_{-,i} - \sum_{i=dn} n_{+,i}}{\sum_{i=dn} n_{-,i} + \sum_{i=dn} n_{+,i}},
\label{spin_dn}
\end{equation}
where, $n_{+,i}$ and $n_{-,i}$  denote the number  of protons in the $i^{th}$ cycle,
in spin up and down state, respectively.
Please note, that according to above definitions the following relations are valid:
\begin{eqnarray}
\sum_{i=up} n_{+,i} + \sum_{i=dn} n_{+,i} = N_+, 
\nonumber \\ 
\sum_{i=up} n_{-,i} + \sum_{i=dn} n_{-,i} = N_-.
\label{11}
\end{eqnarray} 
The beam can be  effectively depolarized  
e.g. by assigning to the events in spin up cycles the weights $w$
which can be derived from the requirement that 
the numerator of 
Equation~\ref{polar} has to vanish:
\begin{equation}
w\cdot \sum_{i=up} n_{+,i} + \sum_{i=dn} n_{+,i} - (w\cdot \sum_{i=up} n_{-,i} + \sum_{i=dn} n_{-,i}) = 0.
\label{roznica}
\end{equation}
Thus, combining Equations~\ref{spin_up} and~\ref{spin_dn} with Equation~\ref{roznica} we obtain 
the following formula for the value of $w$:
\begin{equation}
w = \frac{P^{\downarrow}}{P^{\uparrow}} \cdot \frac{\sum_{i=dn} n_{-,i} + \sum_{i=dn} n_{+,i}}
                                             {\sum_{i=up} n_{+,i} + \sum_{i=up} n_{-,i}}
  = \frac{P^{\downarrow}}{P^{\uparrow}} \cdot \frac{L^{\downarrow}}{L^{\uparrow}}
  = \frac{P^{\downarrow}}{P^{\uparrow}} \cdot \frac{1}{L_{rel}}
\label{warunek}
\end{equation}
where $L_{rel}:=\frac{L^{\uparrow}}{L^{\downarrow}}$ denotes the relative luminosity 
for the spin up and down cycles.
Taking into account the numerical values of
$P^{\uparrow}=0.658\pm 0.008$,  $P^{\downarrow}=0.702\pm 0.008$, 
and $L_{rel}~=~0.98468~\pm~0.00056$~\cite{czyzyk,czyzyk-phd} one gets $w=1.083$.

The weight $w$,  assigned to events in spin up cycles, does not change the absolute value of  
cross sections, as the same weight has been applied in both: the calculation of the number of events
originating from the $pp\to pp\eta$ reaction and the determination of the luminosity from the 
$pp\to pp$ elastic scattering.

\subsection{Data analysis}

The method applied to the determination of the differential cross sections
follows the procedures described in~\cite{pawel-prc}, 
therefore for any details the reader is reffered to that paper. Here we shall 
only briefly describe the main steps of the data analysis and emphasize the 
differences between the methods used in both studies.  

After particle identification we continued the analysis 
with the depolarization of the experimental data, according 
to the procedure described in Section~\ref{depol}. Next, we determined the covariance matrix 
and performed the kinematical fitting~\cite{hab,pawel-prc}.

\begin{figure}[H]
\begin{center}
  \vspace{-0.8cm}
  \includegraphics[width=4.3cm]{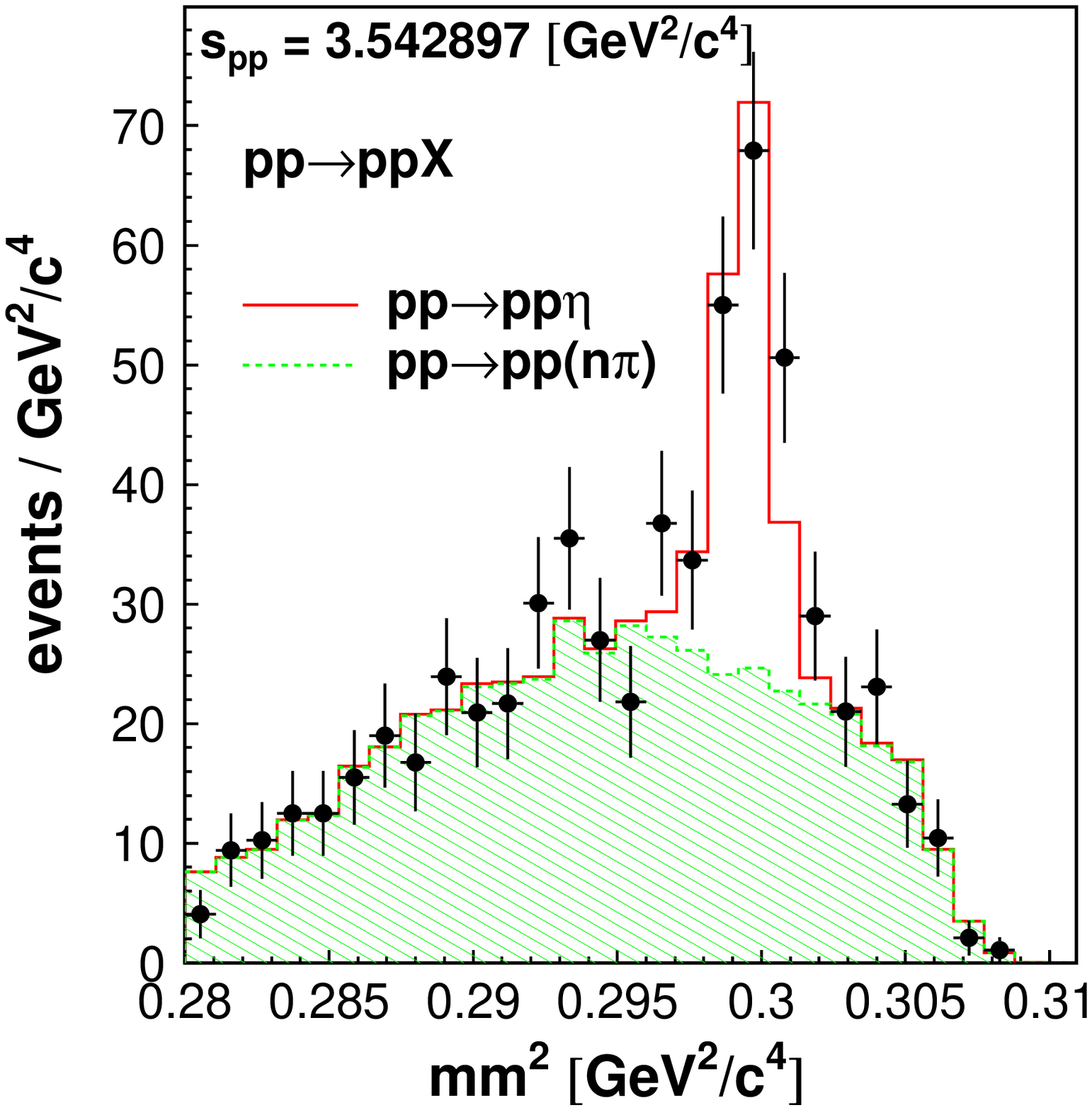} 
  \vspace{-1.0cm}
  \includegraphics[width=4.3cm]{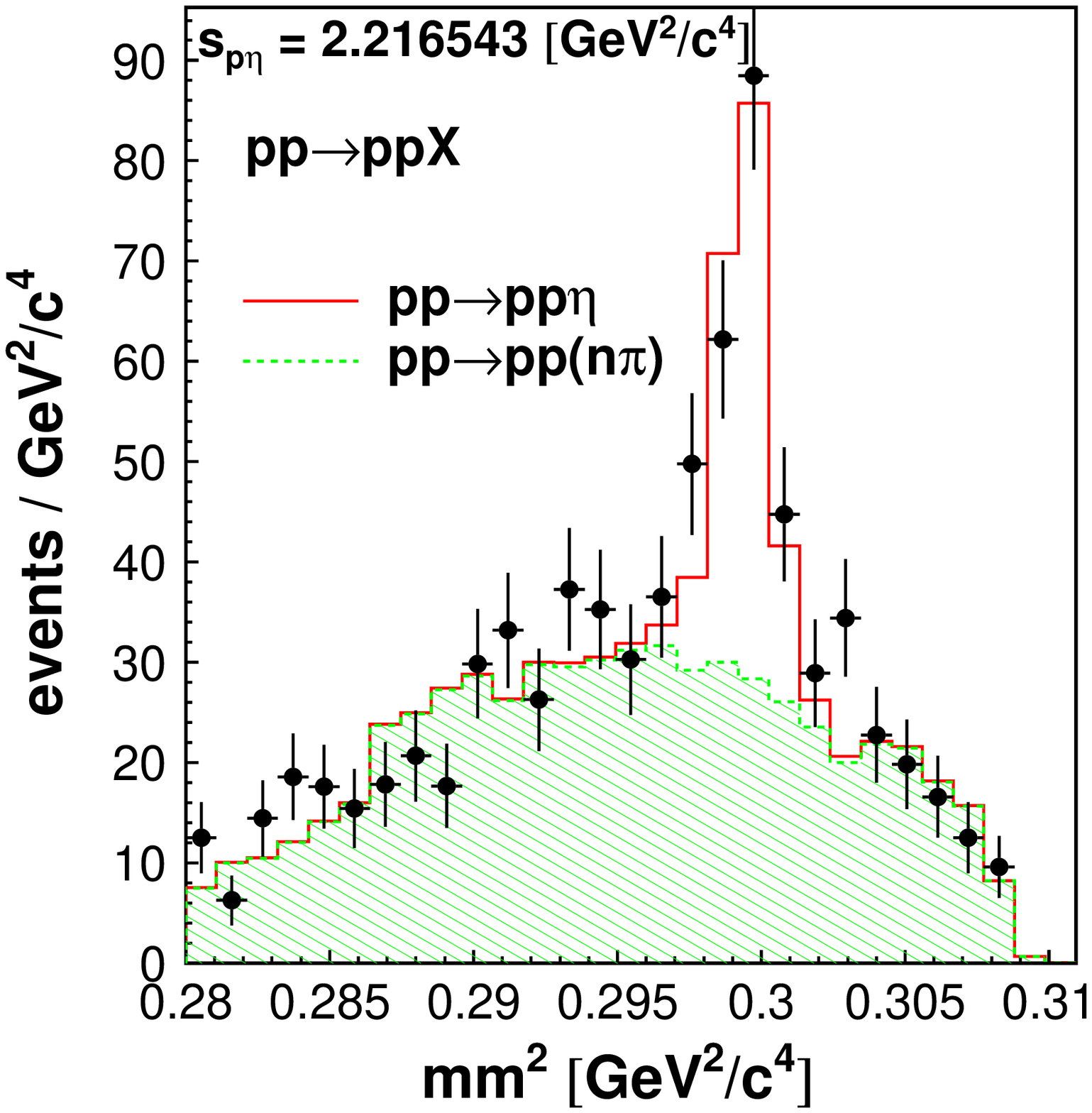} 
  \vspace{-0.3cm}
  \includegraphics[width=4.5cm,height=4.3cm]{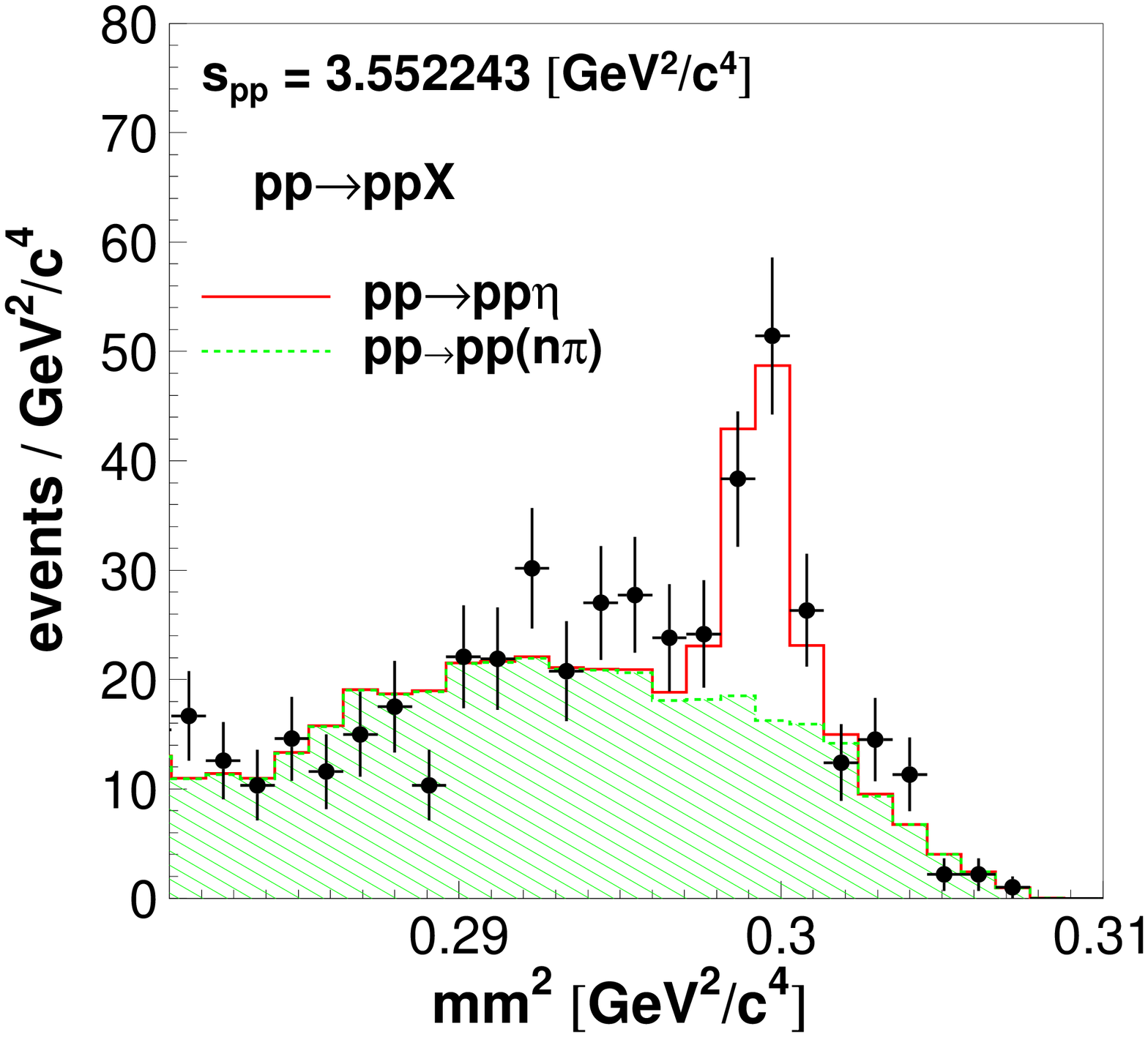} 
  \hspace{-0.45cm}
  \includegraphics[width=4.5cm,height=5.0cm]{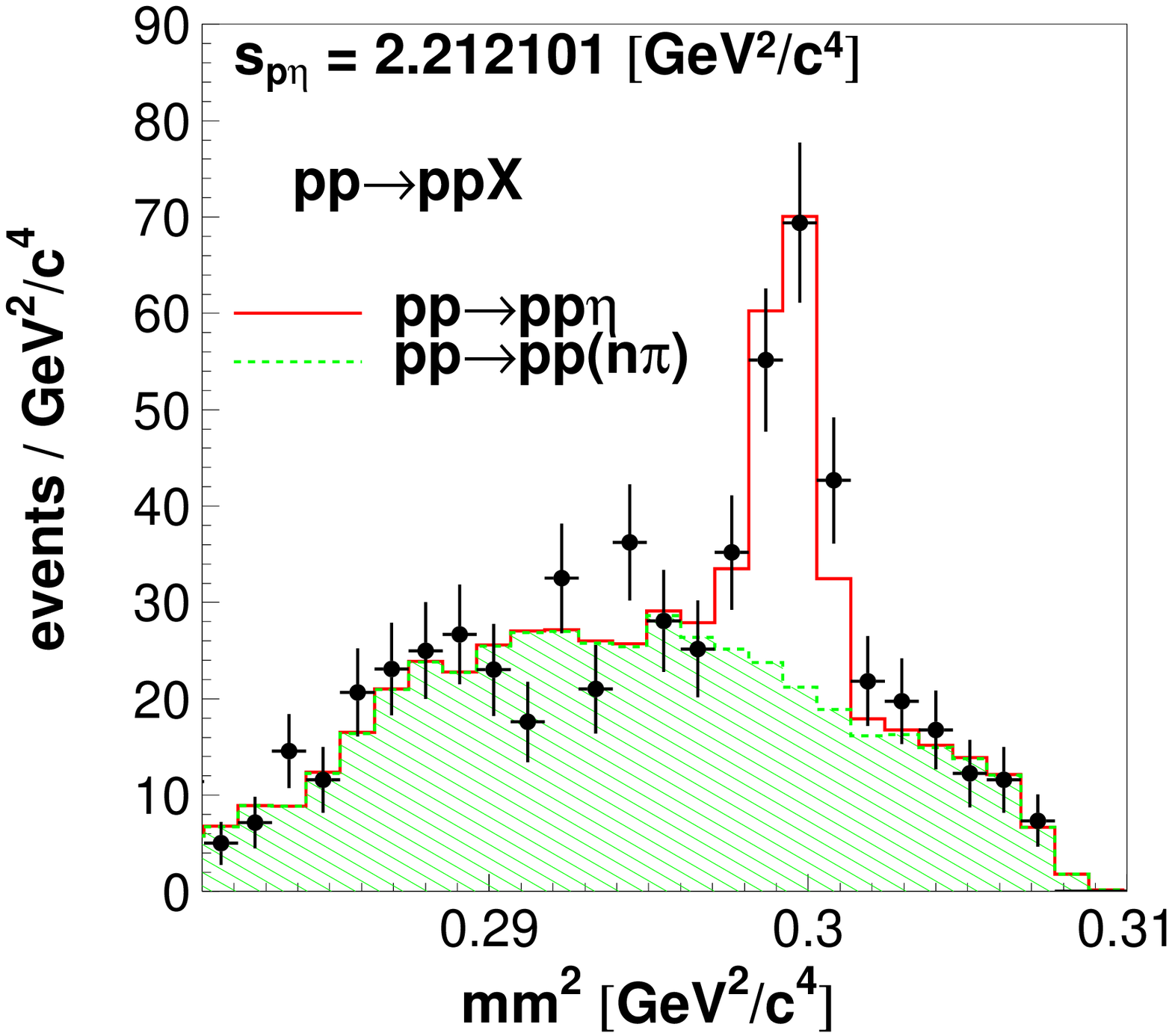} 
  \includegraphics[width=4.5cm,height=4.3cm]{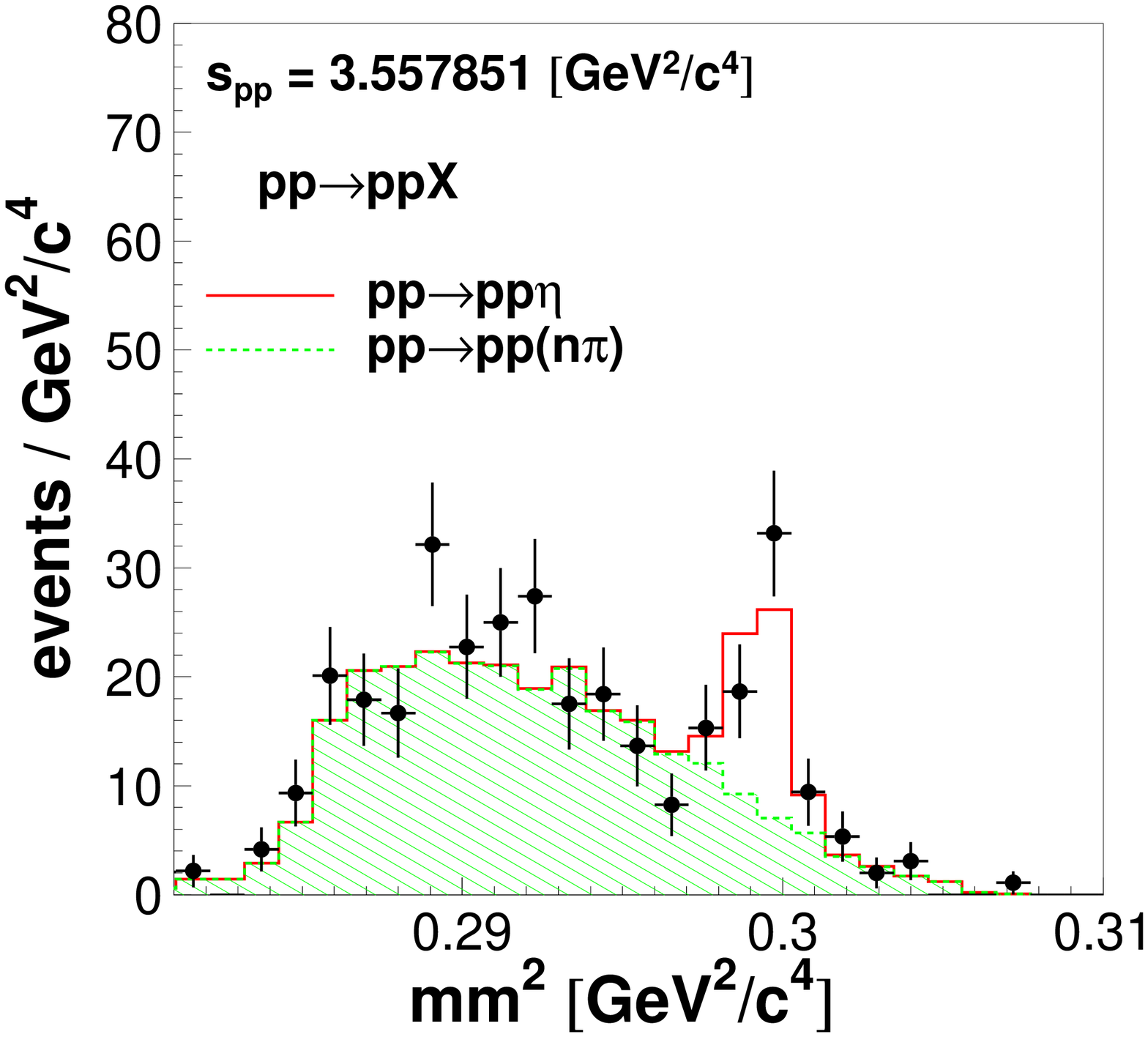} 
  \hspace{-0.45cm}
  \includegraphics[width=4.5cm,height=4.3cm]{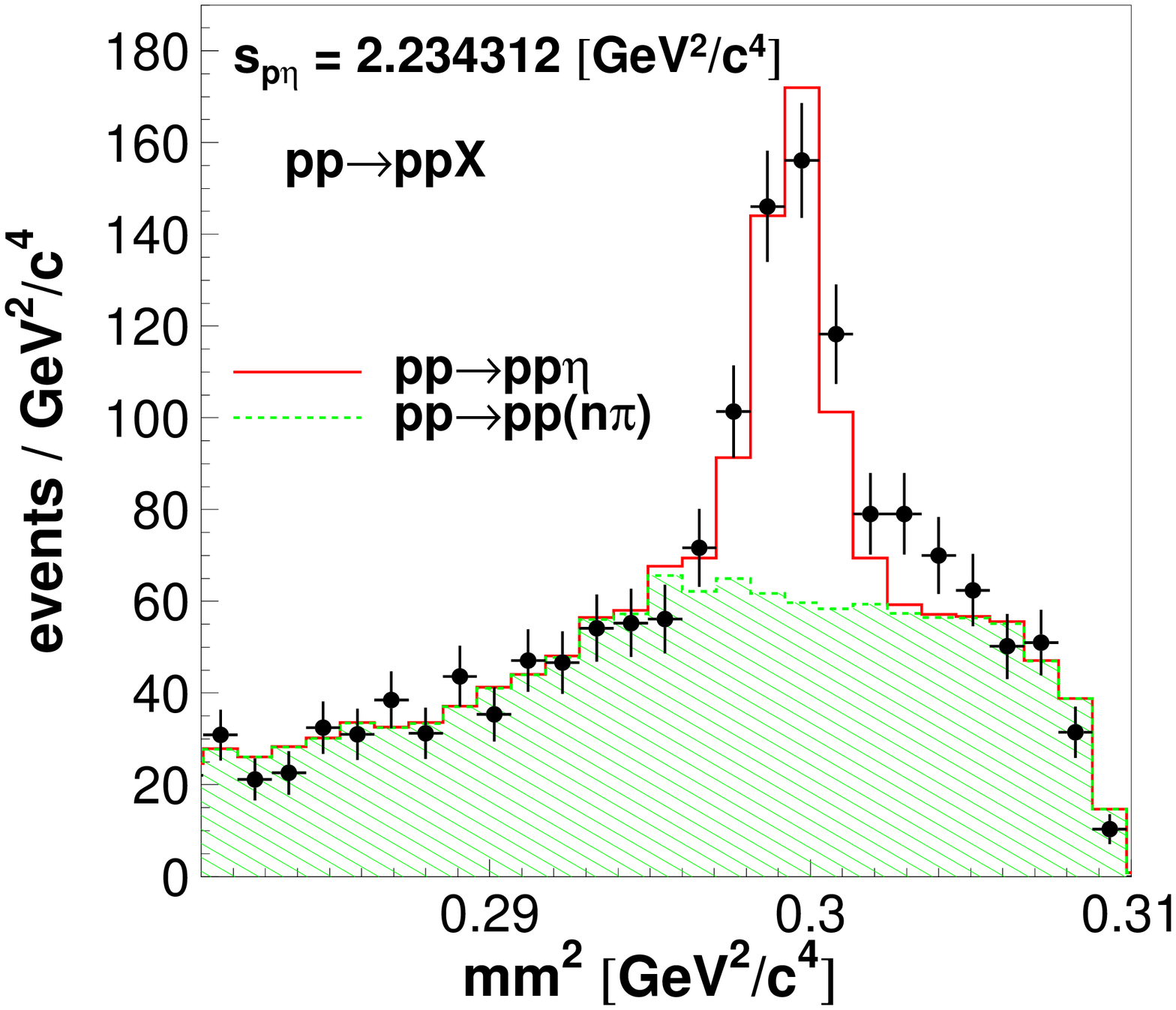} 
\end{center}
\caption{
Missing mass squared distributions for  arbitrarily chosen bins of 
$s_{pp}$ (left) and  $s_{p\eta}$ (right), as measured
at the excess energy of Q~=~10~MeV. Dots represent 
the experimental data points along with their statistical erros, whereas
the solid line is the best fit of the sum of the signal and background 
as obtained in the Monte-Carlo simulations. The shaded part shows the generated
multi-pionic background. 
\label{example}}
\end{figure}

For the description of the relative motion of the protons and the $\eta$ meson,
following reference~\cite{pawel-prc}, we have chosen the 
squares of the invariant masses --~$s_{pp}$ and $s_{p\eta}$~-- 
of the proton-proton and proton-$\eta$ systems, respectively.
Optimizing the statistics we have divided the range of $s_{pp}$ and $s_{p\eta}$ into 20 bins. 
For each bin of these variables 
we have determined the  spectrum of the square of the missing mass.  
Analogoues spectra were simulated for the $pp\to pp\eta$ reaction and for the background channels.
The simulation program,   based on the GEANT3~\cite{geant} code, 
accounts for the geometry of the COSY-11 detector setup
including the conditions of the beam and target~\cite{NIM} and 
the resolution and efficiency of the detectors~\cite{hab}.
The simulated events were analysed with the same program as the experimental data.
Subsequently, 
functions of the type:
\begin{eqnarray}
f(mm^2) = \alpha \cdot f_{pp\to pp2\pi}(mm^2) + 
\beta \cdot f_{pp\to pp3\pi}(mm^2) +
\nonumber \\
+ \, \gamma \cdot f_{pp\to pp4\pi}(mm^2) + 
\delta \cdot f_{pp\to pp\eta}(mm^2), \ \ \ \ \ 
\label{func1}
\end{eqnarray}
were fitted to the data, with  
$\alpha$, $\beta$, $\gamma$, and $\delta$  treated as free parameters
responsible for the normalisation of the simulated missing mass spectra ($f$) 
of reactions indicated in subscripts.
In order to determine the background free invariant mass distributions
the experimental missing mass squared spectra were fitted 
separately for each bin of $s_{pp}$ and  $s_{p\eta}$.
As an example, 
the missing mass distributions
for arbitrarily chosen bins of $s_{pp}$ and
$s_{p\eta}$ are presented in Figure~\ref{example}.
The mumbers of $\eta$ mesons $N_{\eta}$ for the individual intervals of 
$s_{pp}$ and  $s_{p\eta}$
have been calculated as: 
\begin{equation}
N_{\eta} = \delta \cdot \int_{0}^{mm^2_{max}} f_{pp\to pp\eta}(mm^2) \ d(mm^2), 
\label{number}
\end{equation}
and the statistical errors $\sigma(N_{\eta})$ have been estimated as:
\begin{equation}
\sigma(N_{\eta}) = \sigma(\delta) \cdot \int_{0}^{mm^2_{max}} f_{pp\to pp\eta}(mm^2) \ d(mm^2),
\label{number_err}
\end{equation}
where $\sigma(\delta)$ are the estimates of the $\delta$ 
parameter uncertainties (standard deviations) determined by means of the MINUIT 
minimization package~\cite{minuit}. 

The systematic error of $N_{\eta}$ has been estimated to be not larger than 8\%, based 
on the dependence of the results on different assumptions
for i) background estimation ($\sim$2\%),
ii) description of the 
proton-proton final state 
interaction~($\sim$5\%), 
and iii) inaccuracy in efficiency for reconstruction of both proton trajectories ($\sim$6\%).
The uncertainty in the number of the background events
under the peak of the $\eta$ meson 
was estimated as differences  between results obtained
 by fitting to the background a) the first order polynomials,
    b) the  second order polynomials,
    c) the sum of two Gaussian functions,  and
    d) the distributions simulated for the multi-pion
    production~\cite{klajaphd,czyzyk-phd}.
The inaccuracy due to the model 
used for the description of the
proton-proton FSI
was estimated conservatively as  
a difference in results 
determined when using parameterization 
of the proton-proton S-wave interaction~\cite{swave}
and when neglecting the FSI and taking into account
a homogeneous phase-space distribution of the momenta of final state particles.    
%
%
%

The determination of the luminosity was based 
on the comparison of the measured  differential $pp\to pp$ elastic scattering cross sections
to the data of the EDDA group~\cite{edda}. 
For the detailed method of the luminosity and acceptance calculation the interested reader 
is referred to~\cite{hab,NIM}. 
The integrated luminosity was extracted
to be $L=58.53\pm 0.03$~nb$^{-1}$.

\section{Results and conclusions}
\label{results}
The total cross section evaluated 
as an integral of the $s_{pp}$ distribution  equals to $\sigma = 1.27 \pm 0.04 \pm 0.13~\mu$b, 
where the first error is the statistical and the second the  systematic one. 
The latter accounts for the quadratic sum of independent contributions from 
8\% systematic error of $N_{\eta}$,  3\% 
systematic error of the luminosity determination~\cite{pawel-prc}, and 6~\% uncertainty in the acceptance
estimation~\cite{pawel-prc}.
The determined  total cross section
is in line with the previous measurements 
performed independently by various experimental 
groups~\cite{pawel-prc,etadata}.

\begin{figure}[H]
\begin{center}
  \includegraphics[width=4.3cm]{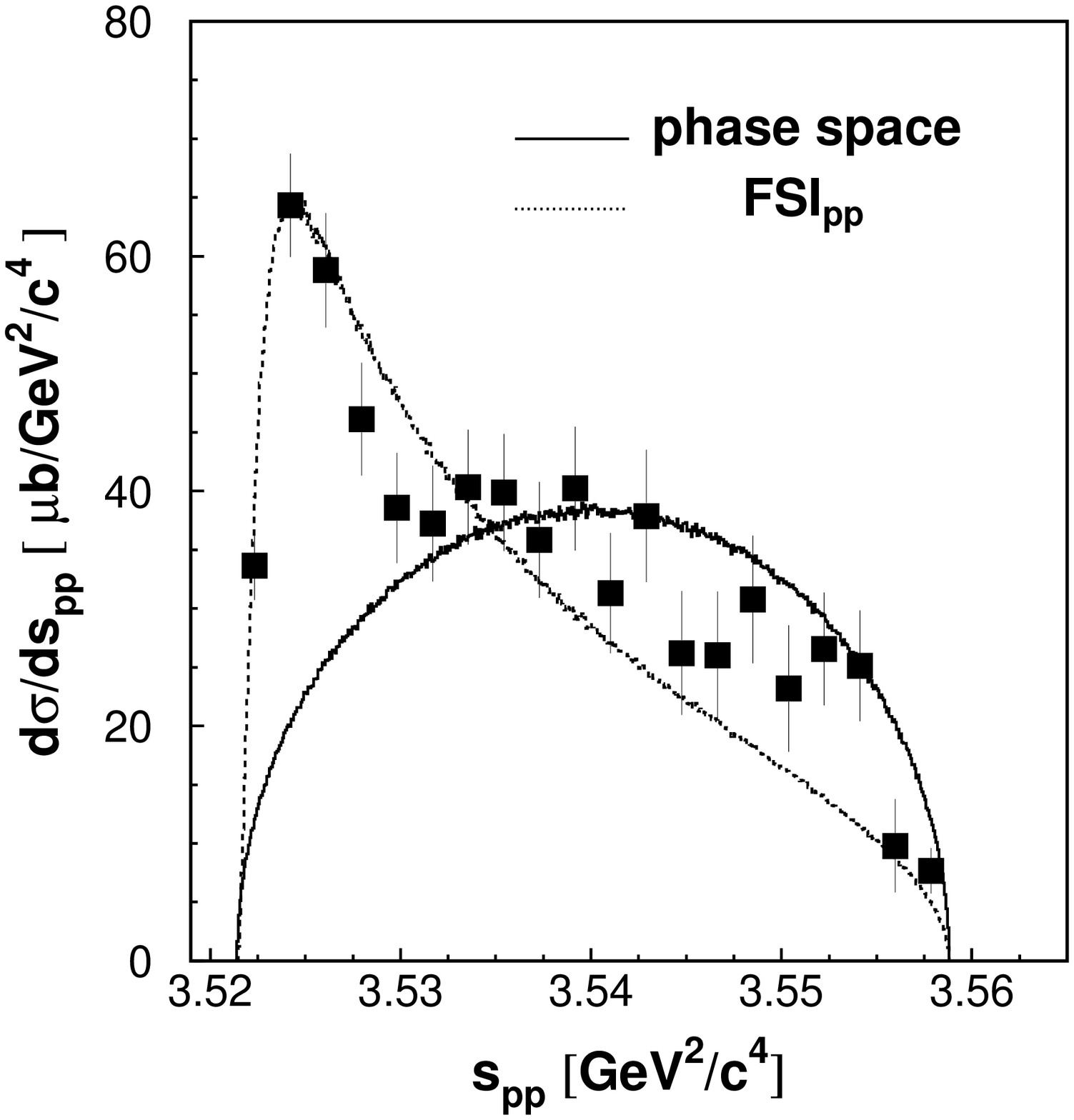}
  \includegraphics[width=4.3cm]{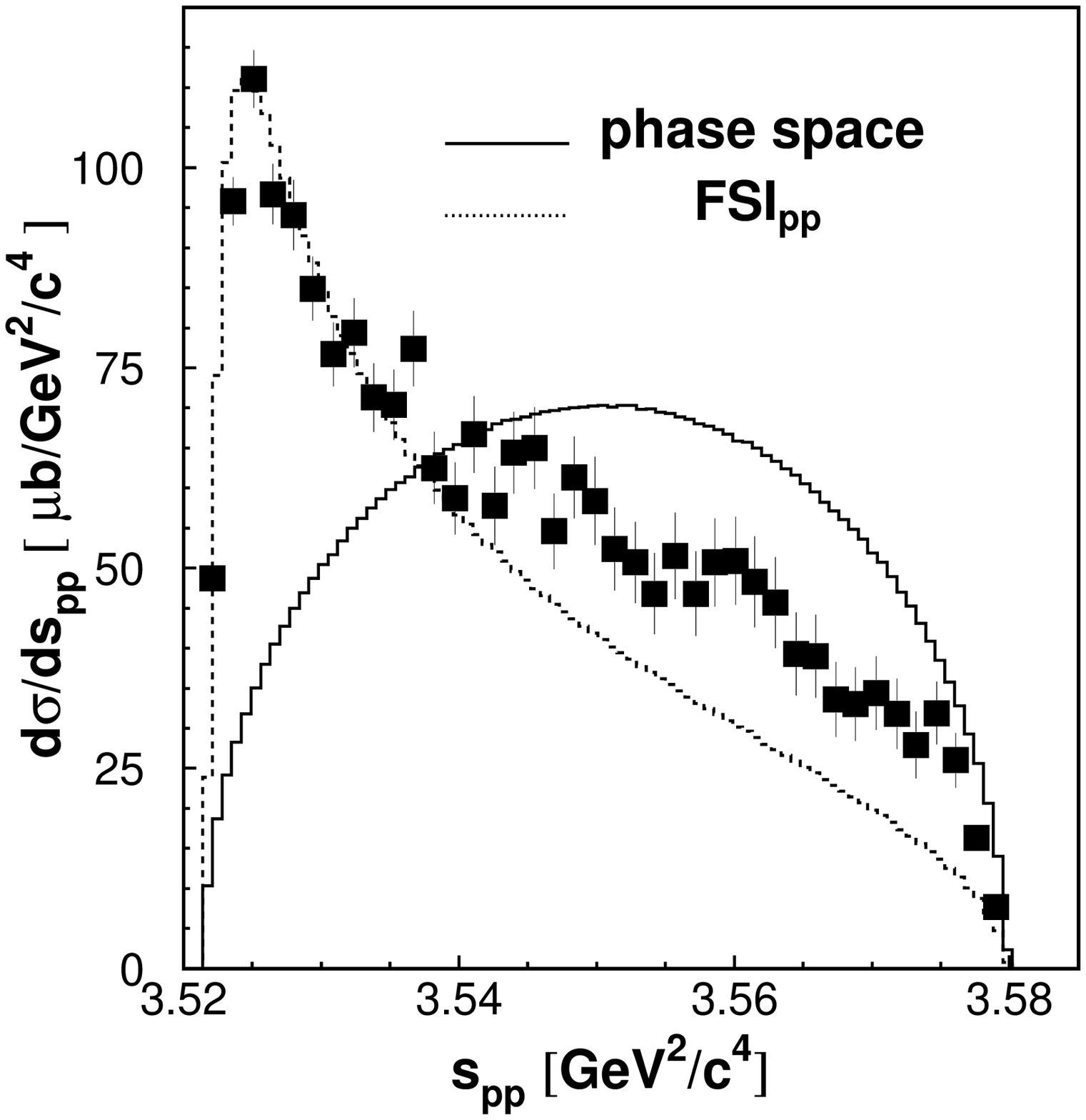}
  \includegraphics[width=4.3cm]{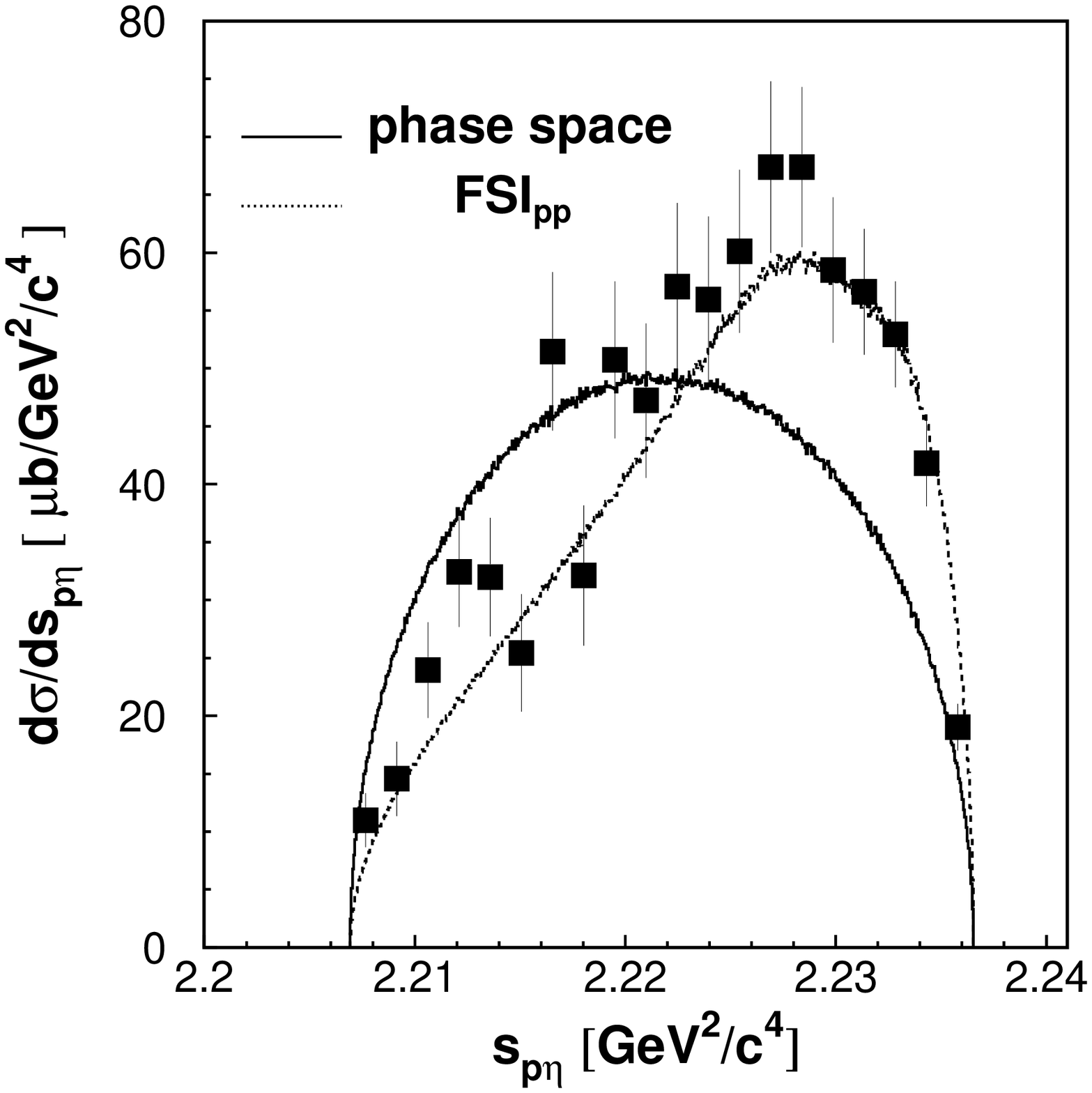}
  \includegraphics[width=4.3cm]{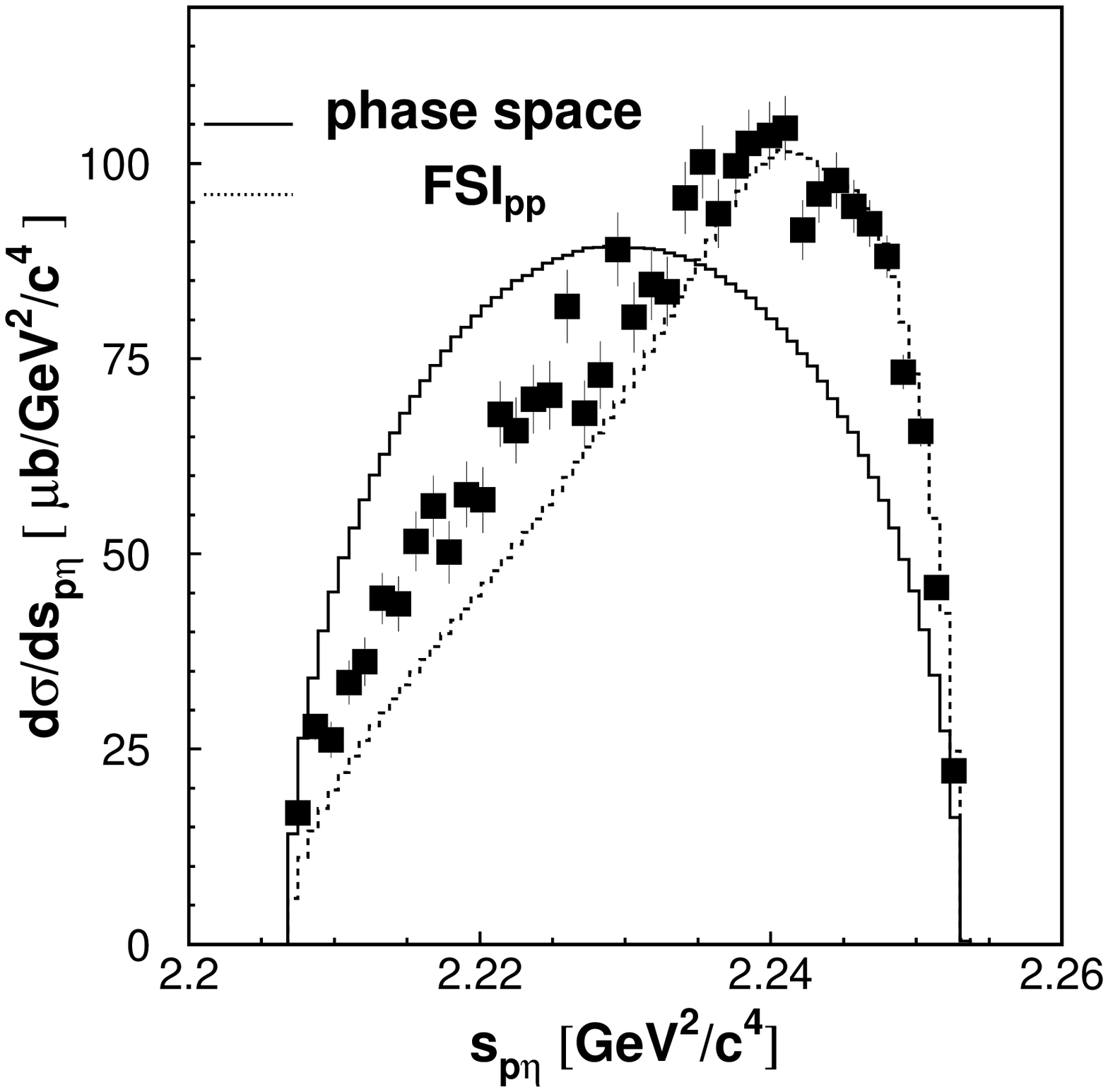}
\end{center}
\caption{ Distributions of the square of the proton-proton and proton-$\eta$ 
invariant masses as measured  at Q~=~10~MeV (left) 
and Q~=~15.5~MeV (right). Solid lines represent the homogeneous phase-space 
distributions, while the dotted lines are the theoretical predictions 
taking into account the $^{1}S_{0}$ proton-proton final state interaction.  
\label{spp}}
\end{figure}

The results on the differential cross sections for the $pp\to pp\eta$
reaction as a function of the square of the proton-proton and 
proton-$\eta$ invariant masses are given in Table~\ref{table1}, 
and presented in the left part of Figure~\ref{spp}. 
They are compared with the differential cross sections measured at the 
excess energy of Q~=~15.5~MeV~\cite{pawel-prc}, displayed in the right part of Figure~\ref{spp}. 
The homogeneous phase-space distributions, shown in Figure~\ref{spp}  
by  solid lines, completely disagree with
the experimental data for all distributions
presented. The peak observed 
at small values of $s_{pp}$ is associated with strong
proton-proton final state interaction (FSI).  
On the other hand the dotted lines, which represent the phase-space 
distribution convoluted with the proton-proton FSI 
describe the data quite well in the range of small invariant masses of the 
proton-proton system and in the range of large $s_{p\eta}$, but for both excess energies
there
is a significant deviation of the theoretical predictions from the experimental 
data in the range of large values of $s_{pp}$ and small values of
$s_{p\eta}$. In the calculations 
we have used the parameterization of the proton-proton FSI given in reference~\cite{baru}. 
The curves including the proton-proton FSI have been arbitrarily normalized 
in the range of small values of $s_{pp}$ and close to the upper limit of the $s_{p\eta}$ distribution.
\begin{table}[h]
\caption{ Distribution of the square of the invariant mass of the 
proton-proton and proton-$\eta$ systems for the $pp\to pp\eta$ reaction 
at the excess energy of Q~=~10~MeV. \label{table1}}
\begin{center}
\begin{tabular}{||c|c||c|c||}
\hline
\hline
$s_{pp}$ [$\frac{GeV^2}{c^4}$]  &  $\frac{d\sigma}{ds_{pp}}$ [$\frac{\mu~b}{GeV^2/c^4}$] &  $s_{p\eta}$ [$\frac{GeV^2}{c^4}$]  &  $\frac{d\sigma}{ds_{p\eta}}$ [$\frac{\mu~b}{GeV^2/c^4}$] \\
\hline
\hline
3.522337 & 33.6 $\pm$ 2.9 & 2.207658 & 11.0 $\pm$ 2.3 \\ 
3.524206 & 64.3 $\pm$ 4.4 & 2.209139 & 14.6 $\pm$ 3.2 \\  
3.526075 & 58.8 $\pm$ 4.9 & 2.210620 & 24.0 $\pm$ 4.1 \\
3.527944 & 46.1 $\pm$ 4.8 & 2.212101 & 32.4 $\pm$ 4.8 \\ 
3.529813 & 38.6 $\pm$ 4.7 & 2.213582 & 32.0 $\pm$ 5.1 \\
3.531682 & 37.2 $\pm$ 4.9 & 2.215062 & 25.4 $\pm$ 5.1 \\
3.533552 & 40.3 $\pm$ 4.9 & 2.216543 & 51.5 $\pm$ 6.8 \\
3.535421 & 39.9 $\pm$ 5.0 & 2.218024 & 32.1 $\pm$ 6.0 \\ 
3.537290 & 35.8 $\pm$ 4.9 & 2.219505 & 50.7 $\pm$ 6.8 \\ 
3.539159 & 40.2 $\pm$ 5.3 & 2.220985 & 47.2 $\pm$ 6.7 \\ 
3.541028 & 31.3 $\pm$ 5.1 & 2.222466 & 57.1 $\pm$ 7.2 \\ 
3.542897 & 37.9 $\pm$ 5.6 & 2.223947 & 55.9 $\pm$ 7.2 \\ 
3.544767 & 26.2 $\pm$ 5.3 & 2.225428 & 60.1 $\pm$ 7.0 \\ 
3.546636 & 26.0 $\pm$ 5.4 & 2.226908 & 67.4 $\pm$ 7.4 \\ 
3.548505 & 30.8 $\pm$ 5.4 & 2.228389 & 67.4 $\pm$ 6.9 \\ 
3.550374 & 23.2 $\pm$ 5.4 & 2.229870 & 58.5 $\pm$ 6.3 \\ 
3.552243 & 26.6 $\pm$ 4.8 & 2.231351 & 56.6 $\pm$ 5.4 \\ 
3.554112 & 25.1 $\pm$ 4.7 & 2.232831 & 52.9 $\pm$ 4.6 \\ 
3.555982 &  9.8 $\pm$ 4.0 & 2.234312 & 41.8 $\pm$ 3.7 \\ 
3.557851 &  7.7 $\pm$ 1.9 & 2.235793 & 19.0 $\pm$ 2.0 \\
\hline
\hline
\end{tabular}
\end{center}
\end{table}

A preliminary result from a comparative analysis of the $pp\eta$ and $pp\eta^{\prime}$ 
system indicates that the observed enhancement is rather not due to the meson-proton 
interaction~\cite{klajaacta,klajaphd}. 
Preliminary results show that the enhancements in the invariant mass distributions
are also present in case of the $pp\to pp\eta^{\prime}$  reaction~\cite{klajaacta,klajaphd}. 
Due to the fact that the interaction between the $\eta^{\prime}$ meson 
and proton is more than an order of magnitude weaker than the one between the $\eta$ meson and 
proton~\cite{swave}, the explanation that the bump  is caused by the proton-$\eta$  final state interactions 
is rather doubtious.  

One plausible explanation for the bumps observed at higher values of $s_{pp}$ and 
lower values of $s_{p\eta}$ 
is the presence of higher partial waves in 
the final state proton-proton system~\cite{kanzo045201}. 
Already the inclusion of the $^{1}S_{0}\to^{3}\!\!P_{0}s$~\cite{objasnienie} 
transition to the production amplitude of the $pp\to pp\eta$ reaction leads to a quite 
well description of the experimental data in the high values of $s_{pp}$ and 
low values of $s_{p\eta}$, leaving unaltered the description of the experimental 
data at low values of $s_{pp}$ and high values of $s_{p\eta}$, dominated mainly 
by the $^{3}P_{0}\to^{1}\!\!S_{0}s$ transition. However, to cope the P-wave contribution
with the flat angular distributions~\cite{pawel-prc,tof}, it is necessary that 
the amplitude  $^{1}D_{2}\to^{3}\!\!P_{2}s$ vanishes or that it interferes destructively  
with  $^{1}S_{0}\to^{3}\!\!P_{0}s$ transition~\cite{deloff}.
Moreover, the model calculations based on a significant P-wave contribution
underestimates the excitation function 
for the $pp\to pp\eta$ reaction below 20~MeV by a factor of about two~\cite{hab,pawel-prc}. 
Although this deficit 
can be overcome when assuming a relatively strong contribution to the production amplitude
from the 
nucleon resonances, such hypothesis cannot be confirmed at the present stage of the inaccuracies 
of the resonance parameters~\cite{kanzo-acta23}.
The amount of the P-wave contribution
should decrease towards threshold but the enhancement 
observed at Q~=~10~MeV is rather of the same order
as the one at Q~=~15.5~MeV, however in view of the present experimental
statistical and systematic inaccuracies the hypothesis 
of the higher partial wave contribution cannot be excluded.

Another explanation for the bumps observed at higher values of $s_{pp}$ 
were put forward by Deloff~\cite{deloff}
who explains the observed spectra by allowing a linear energy dependence of the leading
$^{3}P_{0}\to^{1}\!\!S_{0}s$ partial wave amplitude. Recently also Ceci, {\v{S}varc 
and Zauner~\cite{ceci1,ceci2}
have shown that the excitation function and 
the enhancement in the invariant mass spectra 
can be very well described by the energy dependence of the
production amplitude when the negative interference 
between the $\pi$ and the $\eta$ meson exchange amplitudes is assumed.
However, for the quantitative confirmation of these hypotheses still more precise data 
on the energy dependence of the enhancement in the invariant mass spectra are required.


\section{Acknowledgements} 
The work was partially supported by the
European Co\-mmu\-nity-Research Infrastructure Activity
under the FP6 programme (Hadron Physics,
RII3-CT-2004-506078), by
the Polish Ministry of Science and Higher Education under grants
No. 3240/H03/2006/31  and 1202/DFG/2007/03,
by the German Research Foundation (DFG), 
and by the FFE grants from the Research Center J{\"u}lich.

\end{document}